\documentclass[useAMS,usenatbib]{mn2e} 

\usepackage{amsmath} 
\usepackage{amssymb} 
\usepackage{graphicx} 
\usepackage{txfonts} 
\usepackage{times}

\newcommand{\araa}{ARA\&A} 
\newcommand{\aap}{A\&A} 
\newcommand{\apj}{ApJ} 
\newcommand{\nat}{Nature} 
\newcommand{\apjs}{ApJS} 
\newcommand{\mnras}{MNRAS} 
\newcommand{\pasp}{PASP} 
\newcommand{\aj}{AJ} 
\newcommand{\apjl}{ApJ} 
\newcommand{\pasj}{PASJ}
\newcommand{\procspie}{SPIE Conference}
 
\title[Exploring the infrared/radio correlation at high redshift]
      {Exploring the infrared/radio correlation at high redshift}
      \author[Ibar et al.] 
	 {Edo Ibar,$^{\! 1}$\thanks{e-mail: ibar@roe.ac.uk} 
	   Michele Cirasuolo,$^{\! 1}$ 
	   Rob Ivison,$^{\! 1,2}$ 
	   Philip Best,$^{\! 1}$ 
	   Ian Smail,$^{\!\ 3}$
	   Andy Biggs,$^{\! 2}$ 
	   \and
	   Chris Simpson,$^{\! 4}$
	   Jim Dunlop,$^{\! 1}$  
	   Omar Almaini,$^{\!\ 5}$
           Ross McLure,$^{\! 1}$ 
	   Sebastien Foucaud,$^{\! 5}$ 
	   \and
	   and Steve Rawlings\,$^{\! 6}$
           \vspace*{5mm}\\
	   $^1$ Scottish Universities Physics Alliance,
                Institute for Astronomy, University of Edinburgh,
	        Blackford Hill, Edinburgh EH9 3HJ\\
	   $^2$ UK Astronomy Technology Centre, Royal Observatory,
	        Blackford Hill, Edinburgh EH9 3HJ\\
           $^3$ Institute for Computational Cosmology, Durham University,
                South Road, Durham DH1 3LE\\
	   $^4$ Astrophysics Research Institute, Liverpool John Moores
	        University, Twelve Quays House, Egerton Wharf, Birkenhead
	        CH41 1LD\\
	   $^5$ School of Physics and Astronomy, University of
	        Nottingham, University Park, Nottingham NG7 2RD\\
           $^6$ Astrophysics, Department of Physics, Denys Wilkinson Building,
                Keble Road, Oxford OX1 3RH
		}

\date{Accepted 2008 February 5; Received 2008 January 11; in original
  form 2007 October 9}

\pagerange{\pageref{firstpage}--\pageref{lastpage}} \pubyear{2008}
 
\def\LaTeX{L\kern-.36em\raise.3ex\hbox{a}\kern-.15em 
    T\kern-.1667em\lower.7ex\hbox{E}\kern-.125emX}

\begin{document} 
 
\label{firstpage} 
 
\maketitle 
 
\begin{abstract} 
We have analysed the 24-$\mu$m properties of a radio-selected sample
in the Subaru-{\it XMM/Newton} Deep Field in order to explore the
behaviour of the FIR/radio relation at high redshifts. Statistically,
the correlation is described by $q_{24}$, the ratio between the
observed flux densities at 24\,$\mu$m and 1.4\,GHz, respectively.
Using 24-$\mu$m data results in considerably more scatter in the
correlation than previous work using data at 60--70\,$\mu$m.
Nevertheless, we do observe a steady correlation as a function of
redshift, up to $z\approx3.5$, suggesting its validity back to
primeval times. We find $q_{24}=0.30\pm0.56$ for the observed and 
$q_{24}=0.71\pm0.47$ for the $k$-corrected radio sample, based on
sources with $\rm 300\, \mu Jy < S_{\rm 1.4GHz} < \rm 3.2\, mJy$ and
$\rm 24\,\mu m$ detections. 
A suitable $k$-correction given by a M82-like mid-IR template suggests 
no extreme silicate absorption in the bulk of our radio sample.
Using thresholds in $q_{24}$ to identify
radio-excess sources, we have been
able to characterise the transition from radio-loud AGN to
star-forming galaxies and radio-quiet AGN at faint ($\lesssim1\,\rm
mJy$) radio flux densities. Our results are in broad agreement with
previous studies which show a dominant radio-loud AGN population at
$>1\,\rm mJy$. The rest-frame $U-B$ colours of the expected
radio-excess population have redder distribution than those that
follow the correlation. This is therefore a promising way to select
obscured Type-2 AGN, with a radio loud nature, missed by deep X-ray
observations. Spectroscopic follow-up of these sources is required to
fully test this method. 
\end{abstract} 
 
\begin{keywords} 
starburst -- galaxies:
active -- galaxies:
high-redshift -- galaxies:
classification, colours -- galaxies.
\end{keywords} 
 
\section{Introduction} 

The far-infrared/radio (FIR/radio) relation is one of the tightest
correlations known in astronomy. Established over three decades ago
(e.g.\ Van der Kruit \citealt{VanDer71}; \ Helou et al.\
\citealt{Helou85}), it covers about five orders of magnitude
in luminosity (Condon \citealt{Condon92}; Garrett
\citealt{Garret02}). The correlation is tightest for local
star-forming galaxies 
and relates the integrated thermal emission, from the dust present in
star-forming regions, with the non-thermal synchrotron emission
produced by relativistic particles (cosmic rays) accelerated by
supernovae explosions (Harwit \& Pacini
\citealt{Harwit75}). A complete explanation of the origin of this
remarkable correlation is essential for our understanding of how
galaxies formed and evolved, particularly if the correlation extends
from primeval galaxies to the fully mature galaxies observed locally,
as recent studies have hinted (Kov\'acs et al.\ \citealt{Kovacs06}).

Being relatively direct probes of recent star formation, FIR and radio
data are amongst the most powerful tools available to test whether
primordial galaxies shared the same properties as galaxies present
today. Low-frequency radio data has been used to estimate redshifts using
radio-to-submm flux density ratios (e.g.\ Carilli \& Yun
\citealt{Carilli00}), and to determine the cosmic star
formation history (e.g.\ Haarsma \citealt{Haarsma00}; Ivison et al.\
\citealt{Ivison07a}), facilitating the investigation of the cosmic IR
background (CIRB). 
If the relation between flux densities observed in the FIR and radio
wavebands holds at high redshifts, then we can exploit ultra-deep
high-resolution radio imaging to resolve the CIRB and to explore the
properties of luminous distant galaxies -- for example, studies of submm
(or ``SCUBA'') galaxies (Smail et al.\ \citealt{Smail97}; Hughes et
al.\ \citealt{Hughes98}) which are encumbered by the confusion
encountered in the submm (rest-frame FIR) waveband, and studies of
active galactic nuclei (AGN; Beelen et al. \citealt{Beelen06}) that
are subject to selection biases caused by obscuration at shorter
wavelengths. 

While the combination of {\it Spitzer} observations at 24 and
70\,$\mu$m and radio observations at 1.4\,GHz support the universality
of the FIR/radio correlation up to a redshift of approximately unity (Appleton
et al.\ \citealt{Appleton04}), extending this correlation to
higher redshifts requires extremely deep and complementary
observations at radio and IR wavelengths, as well as reliable
estimates of redshift. The work presented here is based on a deep
multi-wavelength study in the Subaru-{\em XMM/Newton} Deep Field
(SXDF). We have exploited 1.4-GHz images from the Very Large Array
(VLA), reaching a 5$\sigma$ flux density threshold of $\sim\rm 35\,\mu
Jy$ and covering an area of approximately 1\,deg$^2$. The region has
also been observed to impressive depths at optical (SuprimeCAM;
Furusawa et al.\ in preparation), near-IR (UKIDSS-UDS; Lawrence et
al.\ \citealt{Lawrence07}) and mid-IR ({\it{Spitzer}} Wide-area IR
Extragalactic Survey -- SWIRE; Lonsdale et al.\ \citealt{Lonsdale03})
wavelengths.

Starburst galaxies, radio-quiet and radio-loud AGN are the three main
populations which can be observed to high redshifts with the data
employed in this work. It is known that radio-quiet AGN (i.e.\ Seyfert
galaxies) follow the FIR/radio correlation (Roy et al.\
\citealt{Roy98}), but 
radio-loud AGN may vary
considerably in radio luminosities due to their extra core and jet
radio emission, thus weakening and biasing the correlation. In theory,
it is possible to identify these radio-loud AGN via their deviation
from the expected 
relation and therefore to determine the fraction of these so-called
``radio-excess'' sources as a function of radio flux density. This would 
then allow us to constrain the predictions of synthesis population models 
for the transition at sub-mJy radio fluxes from radio-loud AGN to
starburst/radio-quiet AGN (Dunlop \& Peacock \citealt{Dunlop90};
Jarvis \& Rawlings \citealt{Jarvis04}). 

The aim of this work is to explore and discuss the correlation between
the monochromatic fluxes at ${\rm 24\mu m}$ and ${\rm 1.4GHz}$ across
a broad range of redshifts and test the capabilities of the relation
for selecting radio-loud AGN (e.g.. Donley et al.\
\citealt{Donley04}). 

In Section~\ref{observations}, we present the multi-wavelength data
available for the SXDF field and used in this
work. Section~\ref{irrcorr} shows the capabilities of the data to 
extend the ${\rm 24\mu m/1.4GHz}$ correlation to higher redshifts and
Section~\ref{evidence_agn} describes the process of detecting
radio-loud AGN at sub-mJy 1.4-GHz flux densities. The discussion and
conclusions are presented in Section~\ref{discussion} and
\ref{conclusion} respectively. Throughout this paper we have used
$\Omega_m=\rm 0.3$, $\Omega_\Lambda=\rm 0.7$ and $H_0=\rm
70\,km\,s^{-1}\,Mpc^{-1}$ (Spergel et al. \citealt{Spergel07}).
 
\section{SXDF observations} 
\label{observations} 

\subsection{Radio observations at 1.4\,GHz} 
 
Radio observations of the SXDF were carried out by Ivison et al.\
(\citealt{Ivison07b}) during July 2003 using the VLA at 1.4\,GHz. The
total integration time was approximately 60 hours, reaching an r.m.s.\
depth of 7\,$\mu$Jy\,beam$^{-1}$ near the centre of the field. 
The radio map is centred at R.A.\ 2h 18m, Dec.\ $-$5$^{\circ}$ 0$'$,
with a synthesised beam of 1.86\,arcsec $\times$ 1.61\,arcsec at
position angle 15$^{\circ}$. The source detection was carried out
using a 5$\sigma$ threshold, sufficient to avoid significant
contamination by spurious sources, obtaining a total number of 563
radio sources in an area of $\sim$0.45\,deg$^2$. 

In order to exploit the larger area covered by previous
multi-wavelength observations in the SXDF, we have also used the
1.4-GHz catalogue presented by Simpson et al.\
(\citealt{Simpson06}). Their observations comprised thirteen VLA 
pointings reaching a mosaic r.m.s.\ noise of
$\sim$12--20\,$\mu$Jy\,beam$^{-1}$. The Simpson et al.\ catalogue 
contains 505 sources and covers 0.8\,deg$^2$ to a flux density limit
of 100\,$\mu$Jy.

By combining both radio samples, we obtained a total sample of 828
radio sources. Extended sources and obvious doubles, identified by
eye, were considered as single emitters. The distribution of radio
flux densities peaks at about 100\,$\mu$Jy and reaches flux densities
as faint as 35\,$\mu$Jy (5$\sigma$) and as bright as 80\,mJy.

\subsection{{\em Spitzer} observations}
\label{spitzer} 
 
The SXDF was observed by {\it Spitzer} as part of the {\it{Spitzer}}
Wide-area Infrared Extragalactic Survey (SWIRE; Lonsdale et al.\
\citealt{Lonsdale03}) Legacy programme. The SXDF is included in 
one of the seven high-latitude SWIRE fields, the {\em XMM}-LSS field,
with coverage of 9.1\,deg$^2$. Observations were obtained at 3.6, 4.5,
5.8 and 8.0\,$\mu$m with IRAC and at 24, 70 and 160\,$\mu$m with MIPS
(Surace et al.\ \citealt{Surace05}).  In order to explore the
FIR/radio correlation, we have used only the detections at
24\,$\mu$m. It is known that use of the monochromatic 24-$\mu$m flux
densities yields a larger scatter than the 70-$\mu$m data (Appleton et
al.\ \citealt{Appleton04}); however, we are limited by the relatively
shallow coverage at 70\,$\mu$m and the more severe confusion at that
wavelength. 

To explore the FIR/radio relation as a function of redshift, we use
the combined radio source catalogue described in \S2.1 to cross-match
it with the conservative ($10\sigma$) SWIRE source catalogue retrieved
from the {\em Spitzer} Centre Archive. We obtained 370 counterparts
with fluxes $S_{\rm 24\mu m}\gtrsim\rm 0.4\,mJy$ using a
12.5-arcsec-diameter aperture for the photometry.

We then extended the catalogue to fainter 24-$\mu$m flux densities by
measuring the observed flux density on the 24-$\mu$m map, using the
same diameter aperture centred at the radio positions. This allowed
the inclusion of 125 sources with 24-$\mu$m flux densities of $ S_{\rm
24\mu m}\gtrsim\rm 200\,\mu Jy$ ($\sim$4\,$\sigma$; Shupe et al., 
\citealt{Shupe07}). In this paper, for all the remaining radio sources
we used upper limits at $S_{\rm 24\mu m}=\rm 200\,\mu Jy$.

\subsection{Subaru and UKIRT Observations} 

The SXDF is home to the very deep optical survey undertaken by
Subaru/SuprimeCAM (Miyazaki et al.\ \citealt{Miyazaki02}). This
comprises five overlapping pointings, providing broad-band photometry
in the $BVRi'z'$ filters to typical 5-$\sigma$ (AB magnitude) depths
of $B=\rm 27.5$, $V=\rm 26.7$, $R=\rm 27.0$, $i'=\rm 26.8$ and $z'=\rm
25.9$, respectively (2-arcsec-diameter apertures). The seeing in the
composite images is $\sim$0.8\,arcsec (Sekiguki et al.\
\citealt{Sekiguki05}, Furusawa et al.\ \citealt{Furusawa08}).
 
The central region of the SXDF is also being observed at near-IR
wavelengths with the Wide-Field Camera (WFCAM) on the 3.8-m United
Kingdom Infrared Telescope (UKIRT). As the deepest tier of the UKIRT
Infrared Deep Sky Survey (UKIDSS, Lawrence et al.\
\citealt{Lawrence07}), the Ultra Deep Survey (UDS) aims to cover
0.8\,deg$^2$ to $K_{\rm AB}=\rm 25$. In this work, we use the UKIDSS
First Data Release (DR1; Warren et al.\ \citealt{Warren07}) where the
5-$\sigma$ point source depths (in AB magnitudes) at $J$ and $K$ are
23.61 and 23.55, with seeing of 0.86 and 0.76\,arcsec, respectively.

We have used the $K$-band as the reference to cross-match with all
other near-IR and optical images. Unfortunately, the UDS map
misses part of the radio coverage of the SXDF field, limiting the
radio sample to an area with 726 radio sources, out of which 639 (88
per cent) have clear counterparts in the UDS $K$ data within a search
radius of 1.5\,arcsec.  

\subsection{Photometric redshifts} 

We obtained reliable photometric redshifts for 586 of the radio
sources detected at near-IR wavebands. The estimation was carried out
using all available detections from the $B,V,R,i',z',J,K$, 3.6-$\mu$m
and 4.5-$\mu$m photometric bands and considering a conservative
criteria: not including either those sources with poor $\chi^2$
fitting nor those contaminated by halos from nearby stars.

The spectral energy distribution (SED) fitting procedure provides a
photometric estimation consistent within $\Delta z/(1+z) =
0.05\pm0.04$ for the UDS galaxies in general (see Cirasuolo et
al.\ \citealt{Cirasuolo07} for details), and spectroscopic redshifts
available for 63 of the 586 radio sources (Yamada 
et al.\ \citealt{Yamada05}, Simpson et al.\ \citealt{Simpson06} 
and the NASA Extragalactic Database) indicate that the photometric
redshifts of the radio sources are also accurate at this level.
The precision of this estimation is good enough to allow an adequate
$k$-correction to the data (see Figure~\ref{q24_z}). The
resulting redshift distribution of the radio sources peaks at about
$z\sim\rm 0.7$ and includes galaxies up to $z\approx3.5$.

Note that since the photometric redshift estimation depends on
the K-band detection, the sample is basically mass limited. Dwarf
galaxies with faint radio emission may be missed in the K-band,
but massive ones are expected to be detected up to very high
redshift. Heavily obscured galaxies, however, are not drastically
affected.

\section{The FIR/radio relation}
\label{irrcorr}

Based on the analysis presented in Section~\ref{observations}, 369
sources (out of potentially 726 due to the missed UDS area) have both
a 24-$\mu$m detection with flux density $S_{\rm 24\mu m}>\rm 200\,\mu
Jy$ and a reliable redshift (49 of which are spectroscopic).

We present in Figure~\ref{q24_z} the FIR/radio relation in terms of
the basic observable parameter, $q_{24}$ -- the ratio between the
monochromatic fluxes at 24\,$\mu$m and 1.4\,GHz, ${\rm log}(S_{24\mu
m}/S_{\rm 1.4 GHz})$ -- as a function of redshift. Ideally, the
correlation should be calculated using the bolometric FIR flux,
however the paucity of long-wavelength data makes this quantity
difficult to estimate prior to the future arrival of SCUBA-2 (Holland
et al. \citealt{Holland06}) and/or the {\em Herschel Space
  Observatory} (Poglitsch et al.\ \citealt{Poglitsch06}).

\subsection{Non $k$-corrected data}

\begin{figure*} 
  \includegraphics[scale=0.62]{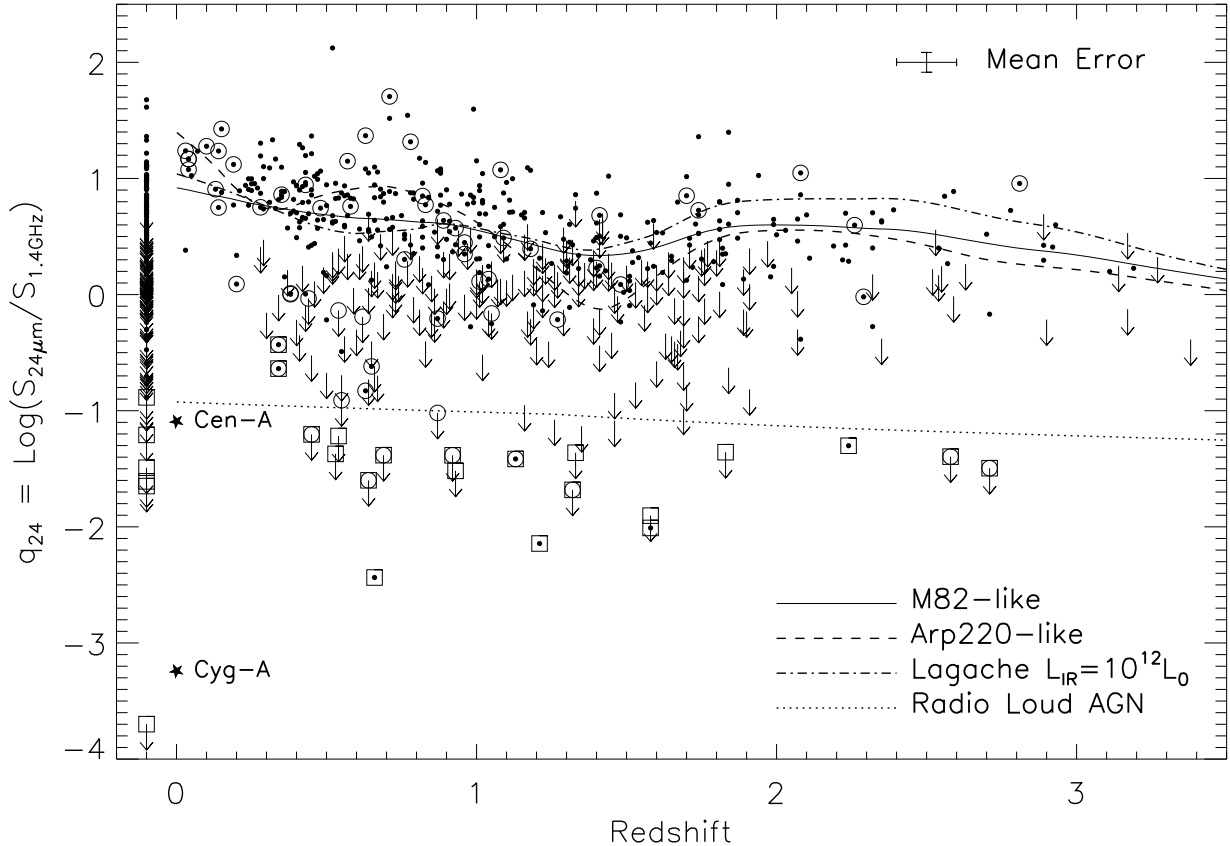}
  \caption{ 
    {\it{Black dots:}} ratio between the observed monochromatic fluxes
    at 1.4\,GHz  and 24\,$\mu$m, $q_{\rm 24}={\rm log}(S_{\rm 24\mu
    m}/S_{\rm 1.4GHz})$, as a function of redshift. {\it{Downward
    arrows:}} upper limits considering a mid-IR flux
    threshold given by 200\,$\mu$Jy. 
    Note that these data have not been $k$-corrected. 
    Sources without photo-z's are plotted at $z=-0.1$.
    {\it{Large circles:}} sources with spectroscopic redshifts.
    {\it{Large squares:}} sources with radio flux densities 
    $S_{\rm 1.4GHz}>\rm 3.2\, mJy$.
    Uncertainties in the observed flux densities give an average
    $q_{24}$  mean error of $\sim0.1$\,dex, and error bars in
    redshift are approximately $\sigma_z\approx\rm 0.1$ (Cirasuolo at
    al.\ \citealt{Cirasuolo07}).
    Overplotted lines show different $k$-correction factors,
    $K(24\mu\text{m})/K(1.4\text{GHz})$, for the observed $q_{24}$
    values as a function of redshift.  The assumed mid-IR templates
    were convolved with the 24-$\mu$m filter from {\it Spitzer}/MIPS
    and are used to show the expected variation of the $q_{24}$ values
    as a function of redshift. Three of these mid-IR templates are
    based on star-forming galaxies: two local standard massive
    starburst galaxies, M82 and Arp\,220, and a SED model for an ULIRG
    galaxy with $L_{\text{IR}}=10^{12}\,\text{L}_\odot$, as described
    by Lagache et al.\ (\citealt{Lagache04}).
    All these three templates are assumed to have a simple spectral
    index $\alpha=-0.7$ for the radio emission, where $S_{\nu}\propto
    \nu^{\alpha}$, and normalised to the best fit to the data
    detections. 
    Finally, a composite spectrum for radio-loud AGN given by Elvis et
    al.\ (\citealt{Elvis94}) is also included, although as indicated
    by the locations of Cen-A and Cygnus-A, radio-loud AGN may be
    distributed over a wide range of $q_{24}$ values. In particular,
    this template 
    is not normalised to the best fit but uses values from the actual
    composite spectrum at both wavelengths.
    }
  \label{q24_z} 
\end{figure*}

Figure~\ref{q24_z} shows that the vast majority of the sources
detected at 24\,$\mu$m follow a tight correlation between $q_{24}$ and
redshift, up to $z\approx\rm 3.5$.  Upper limits have been included
using $S_{\rm 24\mu m}=\rm 200\,\mu Jy$ and show probable evidence for
a large number of radio-excess sources (i.e.\ radio-loud AGN) at the
observed radio flux density regime.

$K$-corrections to the data have been tested assuming standard
starburst galaxy templates for the $24\mu$m detections and a simple
power slope for the radio density fluxes. The expected variations of
$q_{24}$ as a function of redshift are overplotted in
Figure~\ref{q24_z} and are used to study and characterise
the distribution for star-forming galaxies.
Nevertheless, the presence of radio-loud AGN sources in
the sample introduces outliers in the correlation that may cover a
wide range of $q_{24}$ values due to radio excess emission coming
from core and jet structures. For example, the radio-loud AGN
such as 
Cygnus A (a powerful double-lobed radio galaxy) and Centaurus A (one
of the closest known radio galaxies) have $q_{24}=-3.2$ and $-1.1$
values, respectively, far away from the main data trend from
Figure~\ref{q24_z}.  In particular, our data show that the population
with $S_{\rm 1.4GHz}>\rm 3.2\, mJy$ ($\rm 10^{-2.5}\,Jy$; large
squares in Figure~\ref{q24_z}), characterised to be 
AGN-dominated by Condon (\citealt{Condon92}), have $q_{24}\lesssim
-1$, well below the bulk of the data points and similar to the
estimation obtained from a composite radio-loud AGN template (Elvis et
al. \citealt{Elvis94}).
Since we expect and want a correlation valid for star-forming
systems, sources with $S_{\rm 1.4GHz}>\rm 3.2\, mJy$ are excluded
from our analysis of the FIR/radio correlation, basically because these sources
are radio-loud AGN that introduce biases in the statistics.

Using the biweight estimator~\footnote{In order to statistically
describe the data, we used the biweight estimator which is resistant
to outliers and is also robust for a broad range of non-Gaussian
distributions. This is essential to remove remaining AGN from the
statistics of the sample.}
(Beers et al.\ \citealt{Beers90}) to characterise the
data, we find a central location (mean) and scale parameter (sigma; all 
$q_{24}$ values quoted in this paper will similarly be given in this way) 
given by
$q_{24}\rm = 0.66 \pm 0.39$ (based on all detections at 24$\mu$m
including sources without estimated redshifts).  The sub-sample of
sources with $0<z<1$ gives an observed
$q_{24} = 0.80 \pm 0.33$, in agreement with the results of Appleton et
al.\ (\citealt{Appleton04}), i.e.\ $q_{24} = 0.84\pm0.28$.
The data also suggest a decreasing trend of $q_{24}$ as a function of
redshift. A simple linear regression to the $24\,\mu$m-detected
sources with estimated redshift gives a dependency
$q_{24}(z) = (0.85\pm0.01) +(-0.20\pm0.01)\, z$.
Nevertheless, this trend is probably seen because data have not been
$k$-corrected (Hogg et al.\ \citealt{Hogg02}) as we describe later. It
is worth noticing that the available 
49 sources with spectroscopic redshifts support the validity of the
photometric estimation at high redshift (large circles in
Figure~\ref{q24_z}).

\subsection{$k$-corrected data}
\label{k-corrected} 

IR spectroscopic studies of local starburst galaxies have shown three
main components in the mid-IR range (5--38\,$\mu$m): silicate bands
around 10 and 18\,$\mu$m, a large number of PAH emission features, and
a slope of the spectral continuum, where an AGN may be present (Brandl
et al.\ \citealt{Brandl06}). In order to model the $k$-correction to
the mid-IR emission from the bulk of the star-forming sources, we have
convolved the SED of M82 (a local standard starburst galaxy with a
distribution obtained from a fit to the observed photometry presented
by Silva et al.\ \citealt{Silva98}), Arp\,220 (obtained from
photometry listed in the NASA/IPAC Extragalactic Database and compiled
by Pope et al.\ \citealt{Pope06}) and a model of an ultraluminous IR
galaxy (ULIRG; $L_{\text{IR}}=10^{12}\,\text{L}_\odot$) given by
Lagache et al.\ (\citealt{Lagache04}), with the 24-$\mu$m filter
profile from {\it Spitzer}/MIPS. The radio emission is assumed to be
well represented by a power law with a spectral index $\alpha=-0.7$,
based on an average steep-spectrum radio source (Condon
\citealt{Condon92}). 

Note that assumptions of different mid-IR templates
may largely change the $q_{24}$ values at high redshift, mainly 
due to the 10-$\mu$m silicate feature redshifted to $z\approx\rm 1.5$,
where variations of even an order of magnitude in $k$-correction are
seen. On the other hand, if we consider an extreme error in the
radio spectral slope of $\sigma(\alpha)=0.5$ (see findings in Garn et
al. \citealt{Garn07}), we find that the $k$-correction may change in
$\sim0.3$\,dex at $z\approx3.5$, a value much lower than the
24\,$\mu$m uncertainties.

The overplotted lines, in Figure~\ref{q24_z}, show the expected
change in the observed $q_{24}$ parameter as a function of redshift
based on the different adopted templates. Apart from the composite
radio-loud spectra, the $k$-corrections have been normalised to the
best fit to the data.

\begin{figure}
  \includegraphics[scale=0.35]{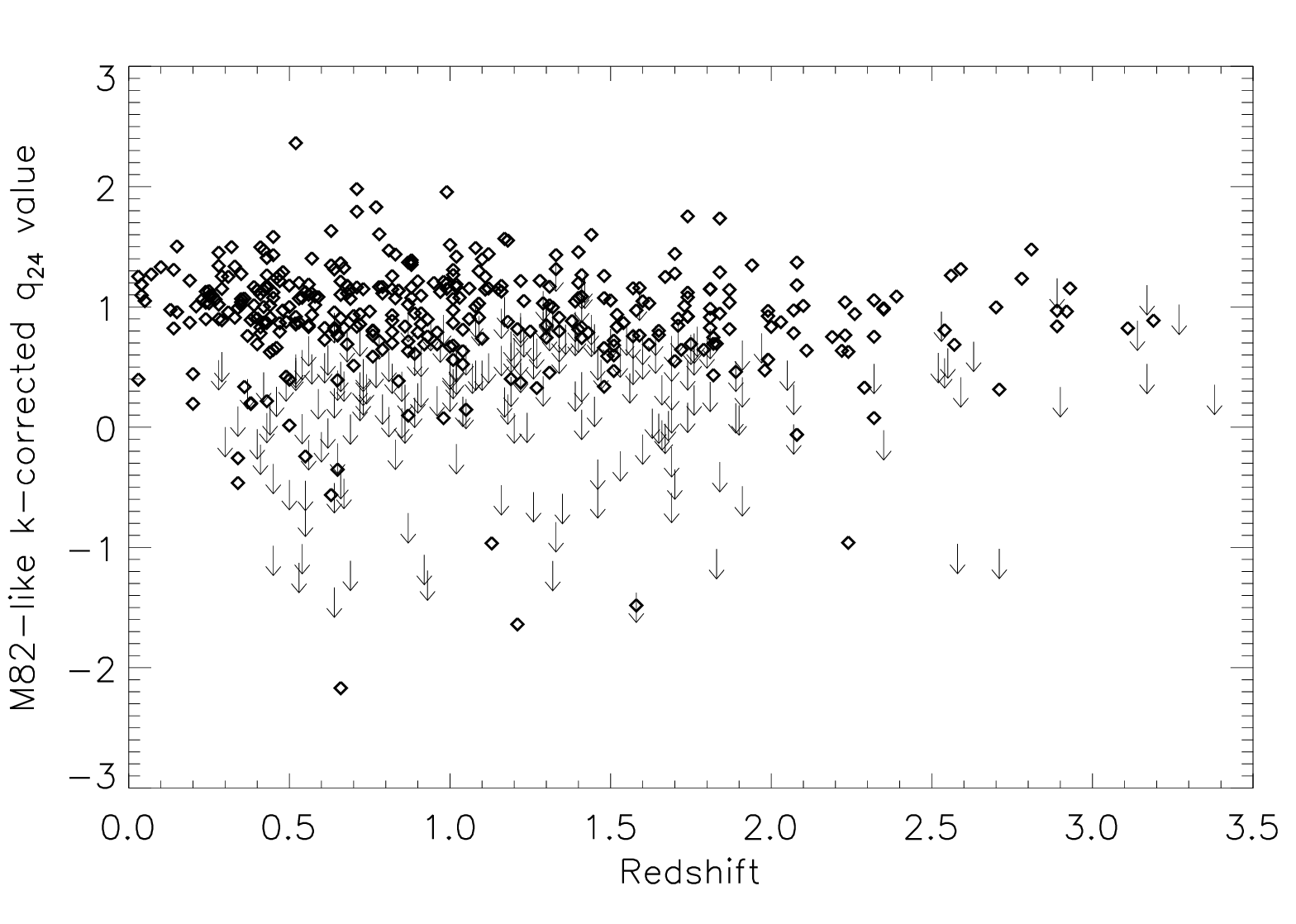}
  \includegraphics[scale=0.35]{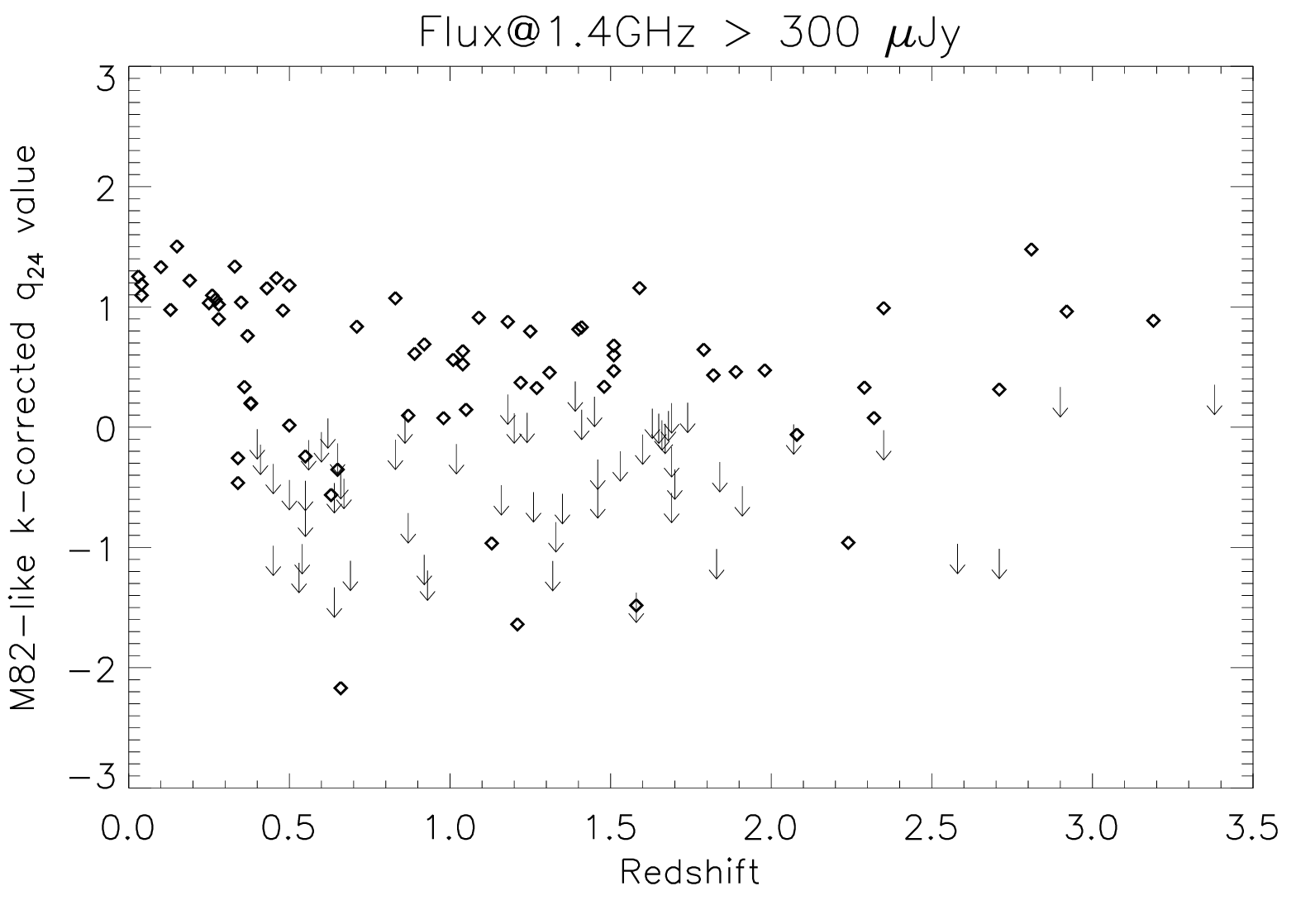}
  \caption{ 
    $K$-corrected $q_{24}$ values as a function of redshift. The
    $k$-correction uses a M82-like template (Silva et al.\
    \citealt{Silva98}) for the mid-IR and a spectral
    slope $\alpha=-0.7$ for the radio emission
    ($S_\nu\propto\,\nu^{\alpha}$). {\it{Top:}} Full 
    radio sample. {\it Bottom:} Radio sources with 1.4-GHz flux densities
    above 300\,$\mu$Jy. 
    Upper limits are considered at $S_{\rm 24\mu m}=\rm 200\,\mu Jy$. 
  }
  \label{q24_z_kcorr} 
\end{figure}

\begin{figure} 
  \includegraphics[scale=0.32]{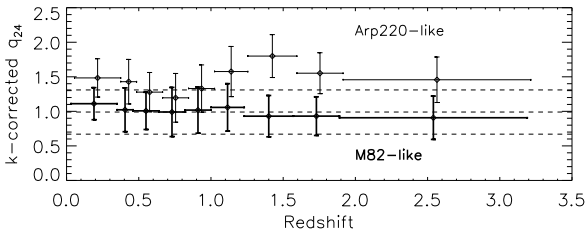}
  \caption{Binned mean and scatter of the M82-like $k$-corrected
    24$\mu$m/1.4GHz correlation (thick black data) as a function of
    redshift, based on the full radio sample.
    The overplotted lines correspond to the mean and one sigma limits,
    for the all-redshift distribution. Also shown (in thin black data)
    are the equivalent values for an Arp220-like $k$-correction
    (shifted in redshift by an small factor in order to improve the
    visualisation only), which results in considerably more residual
    redshift variation. Note that the significantly higher $q_{24}$
    values for the Arp220-like $k$-correction results from the upturn
    in the Arp220-like correction at $z < 0.3$ and due to the
    particularly extreme silicate absorption features
    (cf. Figure~\ref{q24_z}).} 
  \label{evolution} 
\end{figure}

A descending trend in $q_{24}$ as a function of redshift is expected
from all three starburst-like $k$-corrections, in agreement with the
observed slope. The $k$-corrected relation based on sources
with mid-IR detections (${\rm >200\,\mu Jy}$) and
estimated redshifts, changes to 
$q_{24} = 0.99\pm0.32,\ 1.45\pm 0.38,\, $ and $1.08\pm0.37$ (biweight
estimator) using the M82-like, Arp\,220-like and the Lagache et al.\
model, respectively. The M82-like $k$-corrected $q_{24}$ values are
presented in the upper panel of Figure~\ref{q24_z_kcorr} to demonstrate
the improvement in the correlation as a function of redshift.
Figure~\ref{evolution} shows a redshift-binned version of this plot (for
the 24$\mu$m detected sources), demonstrating that both the mean and
scatter of the M82-like $k$-corrected $q_{24}$ value remain constant out
to $z \sim 3.5$; it also demonstrates that $k$-correcting with an
Arp220-like spectrum leads to large variations in the relation with
redshift. 
Based upon the behaviour of the M82-like $k$-corrected distribution,
and particularly the lack of a clear excess in scatter for sources
between $1<z<2$, we suggest no strong MIR-silicate absorption for the
bulk of the sources.

The M82-like $k$-correction to all sources with detections at 24$\mu$m in
the range $0<z<1$ gives $q_{24} = 1.03\pm0.31$, in agreement with previous
Appleton et al.\ (\citealt{Appleton04}) estimation ($q_{24}=1.00\pm0.27$).
A simple linear regression to the M82-like $k$-corrected data is given by
$q_{24}(z) = (0.94\pm0.01) + (-0.01\pm0.01)\ z$, suggesting a constant
dependency (at $1\sigma$ level) for the $q_{24}$ fraction with
redshift. This result assumes that all
galaxies in the sample can be fitted with a single M82-like template,
although the fact that the correlation naturally 
appears to have the same mean and scatter as a function of redshift, up to
$z\sim3.5$, is highly suggestive that the FIR/radio correlation holds all
the way back to primordial times.

\subsection{The correlation based on a complete radio sample at
    24$\mu$m} 

\begin{figure} 
  \begin{center}
  \includegraphics[scale=0.31]{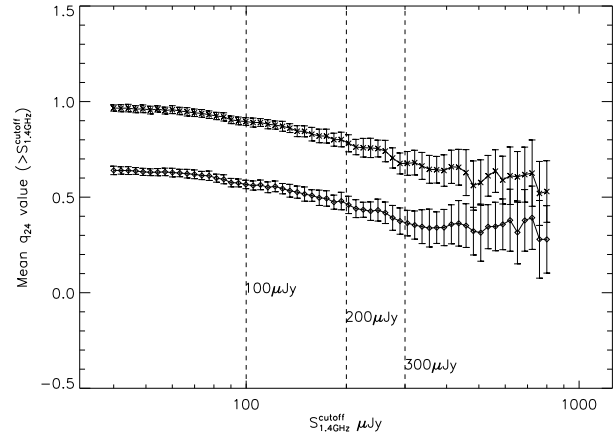}
  \caption{ 
    Mean $q_{24}$ value as a function of cut-off in radio flux
    density. Diamonds and crosses are based on the observed
    and M82-like $k$-corrected data, respectively, both showing
    similar dependencies with the cut-off in radio.
    The error bars represent the uncertainty on the
    mean value estimated using boot-strap re-sample technique.
    The vertical dashed lines are at $100\,\mu$Jy, $200\,\mu$Jy and
    $300\,\mu$Jy. Since sources with 
    $S_{\rm 1.4GHz} >\rm 3.2 mJy$ ($\rm 10^{-2.5}Jy$) are essentially
    all radio-loud AGN (Condon \citealt{Condon84}), these were not
    included in the estimation for the mean value. 
    } 
 \label{mean_cutoff} 
  \end{center}
\end{figure}

The mean and scatter values quoted for the correlations described above 
may be being biased by incompleteness at 24\,$\mu$m for sources
with faint radio fluxes: a large number 
of star-forming dominated radio sources with $S_{\rm 1.4GHz}
\lesssim\rm 300\,\mu Jy$ are not expected to have been detected at
24\,$\mu$m. To demonstrate this problem,
we plot in Figure~\ref{mean_cutoff} the mean $q_{24}$ as a function of
a cut-off in the 1.4-GHz flux density. It clearly shows a decreasing
trend for the mean $q_{24}$ from fainter radio fluxes up to $S_{\rm
1.4GHz} \approx \rm 300\, \mu Jy$, where a flattening starts to
appear. This means that considering a high enough radio cut-off, the
correlation becomes independent of the radio sample. 
We, therefore, base the correlation on a complete radio
sample by constraining the number of sources to those with $S_{\rm
  1.4GHz} >\rm 300\,\mu Jy$. This radio sample is
essentially complete at 24$\mu$m for star-forming galaxies, and all
residual 24$\mu$m non-detection are associated with radio-loud AGN
activity (see below). 

Using the biweight estimator, we find that for the radio population
with $S_{\rm 1.4GHz}>\rm 300\,\mu Jy$ and mid-IR detections
($S_{\rm 24\mu m}>\rm 200\,\mu Jy$), the observed $q_{24}$ values
has significantly lower mean and larger scatter for the
(non-$k$-corrected) correlation
$q_{24} = 0.30\pm0.56$ (it changes to
$q_{24}  = 0.70\pm0.63$ for the sub-sample of sources at
$0<z<1$).

\begin{table*}
    \label{table1}
    \begin{tabular}{llcccc}
      \hline
      &
      Radio sample & 
      $q_{24}$ correlation & 
      Linear regression &
      M82-like $k$-corrected & 
      Linear regression for the \\
      &  
      &  
      mean \& scatter &  
      & 
      $q_{24}$ mean \& scatter & 
      M82-like $k$-corrected $q_{24}$'s 
      \\
      \hline
      (1) &
      $\rm S_{24\mu m}>200\mu Jy$ & 
      $0.66\pm0.39$ & 
      $q_{24}=(0.85\pm0.01)-(0.20\pm0.01)\,z$ & 
      $0.99\pm0.32$ & 
      $q_{24}=(0.94\pm0.01)-(0.01\pm0.01)\,z$ \\
      (2) &
      $\rm S_{24\mu m}>200\mu Jy$ & & & & \\
      & \,\,\&\, \, $0<z<1$ & 
      $0.80\pm0.28$ & 
      &
      $1.03\pm0.31$ & 
      \\
      (3) &
      $\rm S_{1.4GHz}>300\mu Jy$ & & & &\\
      & \,\,\& \,\, $\rm S_{24\mu m}>200\mu Jy$ & 
      $0.30\pm0.56$ &
      $q_{24}=(0.66\pm0.04)-(0.28\pm0.03)\,z$& 
      $0.71\pm0.47$ & 
      $q_{24}=(0.71\pm0.04)-(0.03\pm0.03)\,z$\\
      (4) &
      $\rm S_{1.4GHz}>300\mu Jy$ & & & &\\
      & \,\,\& \,\, $\rm S_{24\mu m}>200\mu Jy$ &  & & & \\
      & \,\,\&\, \, $0<z<1$ & $0.70\pm0.63$ & 
      &
      $0.95\pm0.56$ & 
      \\
      \hline
    \end{tabular}
    \caption{Summary of the results presented in
      Section~\ref{irrcorr}. The linear regressions are based on the
      detected sources at $\rm 1.4GHz$ and $24\mu$m, using intrinsic
      error bars from the source extraction processes.}
\end{table*}

\begin{figure}
  \includegraphics[scale=0.35]{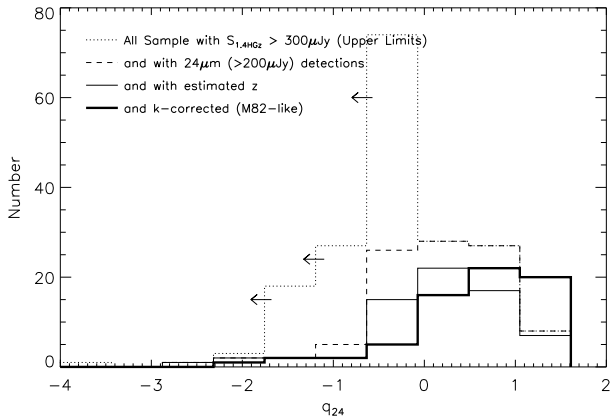}
  \caption{ 
    Histogram of the $q_{24}$ values for the radio sources with
    radio flux densities brighter than $300\,\mu$Jy using different
    selection criteria. Upper limits are assumed considering $S_{\rm
    24\mu m}=\rm 200\,\mu Jy$. The distribution for the $k$-corrected
    data using a M82-like template is plotted by a thick black line.
  } 
  \label{hist_q24} 
\end{figure}

A histogram of $q_{24}$ for the sources with $S_{\rm 1.4GHz}>\rm
300\,\mu Jy$ is shown in Figure~\ref{hist_q24}. It shows that the bulk
of the undetected sources is found below $q_{24}=-0.18$ due to the
200-$\mu$Jy threshold adopted for the mid-IR image. This value of
$q_{24}$ implies sources three times brighter in radio luminosity than
those from our observed mean value
$\langle q_{24}\rangle = 0.30$, and $\sim10.5$ times brighter than
the observed mean value found by Appleton et
al.\ (\citealt{Appleton04}), $\langle q_{24}\rangle = 0.84$. Yun et
al.\ (\citealt{Yun01}) have defined radio-excess sources as those whose
radio luminosity is $\geq5$ times the value predicted from the
FIR/radio correlation. In this sense, radio-loud AGN based on
24-$\mu$m detections are sensitive to the assumed mean value for the
FIR/radio correlation (Appleton et al.\ \citealt{Appleton04}, Donley et
al.\ \citealt{Donley05}, Boyle et al.\ \citealt{Boyle07}). We 
find that 62 per cent of the radio sources (24 detections and 82 upper
limits) plotted in Figure~\ref{hist_q24} have $q_{24}<-0.18$. These
values clearly suggest a large number of radio-loud AGN in this
radio flux regime. 

It is worth noting that the incompleteness produced
by sources without photometric redshifts (compare thick continuum and
dashed lines in Figure~\ref{hist_q24}) does not introduce significant
changes in the mean and scatter of the $q_{24}$ distribution. 

The $k$-corrected mean and scatter values (biweight estimator) for the
distribution of sources with $S_{\rm 1.4GHz}>\rm 300\,\mu Jy$, mid-IR
detections (${\rm >200\,\mu Jy}$) and estimated redshifts, changes to
$q_{24} = 0.71\pm0.47,\ 1.25\pm 0.47,\, $ and $0.77\pm0.48$ after
using the M82-like, Arp\,220-like and Lagache et al.\ model,
respectively. The bottom panel in Figure~\ref{q24_z_kcorr} shows the
M82-like $k$-corrected $q_{24}$ values as a function of redshift for
this sub-sample. It suggests a slight descending trend, from $z<0.5$
to higher redshifts, probably
associated with the presence of steeper mid-IR spectra for
brighter radio sources, although it is not possible to statistically
discriminate a better representative SED between the tested
templates.

A simple linear regression to the M82-like $k$-corrected $q_{24}$
distribution, based on radio sources brighter than 300$\mu$Jy and
detected at 24$\mu$m, is given by $q_{24}(z) = (0.71\pm0.04) +
(-0.03\pm0.03)\ z$. For this complete radio sample, we also find a
flat distribution at $1\sigma$ level, suggesting the validity of the
FIR/radio correlation up to $z\approx3.5$. Note, however that even at
these bright radio flux densities, at $z\gtrsim1$ the upper limits may
overlap with the mean value found for the correlation, suggesting the
possibility of confusion at higher redshifts between radio-loud AGN
and those affected by mid-IR features in the SEDs. Future studies with
submm and FIR facilities, such as SCUBA-2 and the {\em Herschel Space
  Observatory}, will help to better estimate the total bolometric FIR
luminosity, and therefore discriminate between those sources affected
by SED features or due to powerful radio AGN activity.

\section{Radio-loud AGN activity}
\label{evidence_agn} 

\subsection{The transition form radio-loud AGN to star-forming
    dominated galaxies}  

Making use of the M82-like $k$-corrected data presented in the last
section, we have worked out a method of selecting radio-loud AGN
from our radio sample, based on different $q_{24}$ thresholds.

It is well known that most powerful radio sources ($>1$\,mJy) are
radio-loud AGN (Condon \citealt{Condon92}), but the number
counts obtained from deep radio surveys have suggested the appearance
of a new, dominant population at fainter radio flux densities. It is
well established that this faint radio population may be composed of
star-forming galaxies (Condon \citealt{Condon84}) and radio-quiet AGN
(Jarvis \& Rawlings \citealt{Jarvis04}). However, the transition from
radio-loud AGN to star-forming galaxies is still not well defined.

To look at the relative fractions of these populations as a function
of radio flux density, we use the previous M82-like $k$-corrected data to
identify radio-loud AGN based on different $q_{24}$ cut-off thresholds
(Donley et al.\ \citealt{Donley05}). The main assumptions for the
following results are the mean and scatter of the ${\rm 24\,\mu
m/1.4\,GHz}$ correlation (\S 3.3), and the flux thresholds from the
radio and mid-IR maps.

\begin{figure} 
  \includegraphics[scale=0.24]{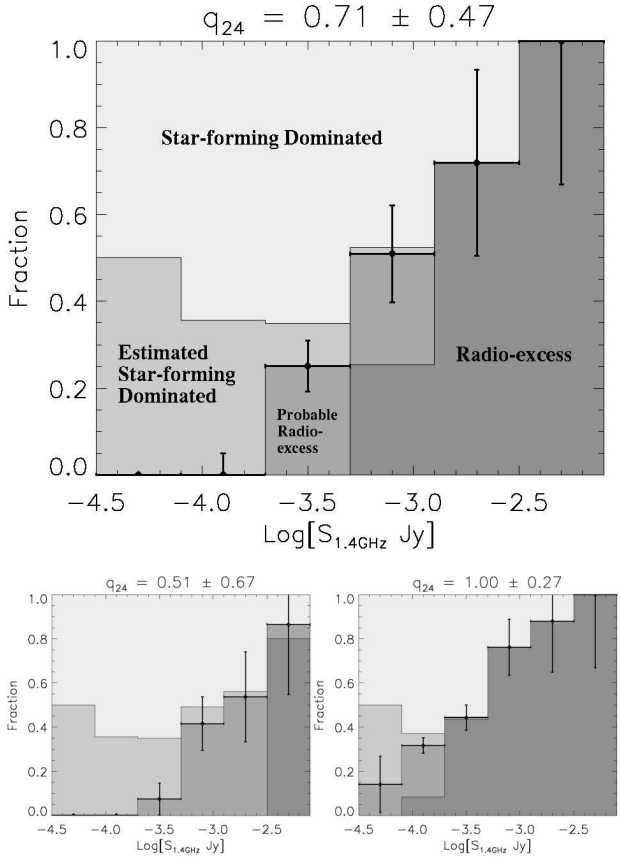}
  \caption{ 
    The sub-mJy radio flux transition from radio-excess to
    starburst-dominated 
    sources based on different $q_{24}$ thresholds using the M82-like
    $k$-corrected data. Sources without redshift have been
    included using a mean M82-like $k$-correction ($+0.33$), non
    varying considerable the results but improving error bars. The
    ``Radio-excess'' fraction is composed of those sources detected at
    24\,$\mu$m, or with upper limits, with 
    $q_{24}$ less than 2\,$\sigma$ from the mean value of the
    correlation. The ``Star-forming Dominated'' fraction is,
    therefore, that with $q_{24}$ values above the 2\,$\sigma$
    threshold for radio-excess selection. The ``Probable Star-forming
    Dominated'' fraction is based on an estimation which corrects the
    incompleteness at 24\,$\mu$m of the ``Star-forming Dominated''
    sources (see text for details). Hence, all the rest of the sources
    are very unlikely  to follow the ${\rm 24\,\mu m/1.4\,GHz}$
    correlation and are called the ``Probable Radio-excess'' fraction.
    Error bars, based on the number counts, are 1-$\sigma$ confidence
    limits given by Gehrels (\citealt{Gehrels86}). The main upper
    panel shows our best 
    estimate of the different populations, based on the results from
    \S3.3. The small panels at the bottom consider the same
    grey-colour criteria. On the left-hand side we modify the $q_{24}$
    mean and scatter based on the higher end of the correlation
    ($\langle q_{24} \rangle+\sigma_{q_{24}}\approx1.2$), and on the
    right-hand side we use the previous Appleton et al.\
    (\citealt{Appleton04}) results. 
  }
  \label{fraction24um} 
\end{figure}

Figure~\ref{fraction24um} shows three different diagrams for the expected
fraction of starburst/radio-quiet AGN (`star-forming-dominated'
systems\,\footnote{The reason we call these objects as
  star-forming-dominated sources comes only from the fact that they 
  follow the FIR/radio correlation. Although, this is not entirely
  true because the radio emission from radio-quiet AGN (luminous
  Seyferts; sources that have been found to follow the correlation)
  takes the form of jets and suffers the same 
  orientation-dependent beaming effects as the radio-loud AGN,
  implying it is most definitely AGN-related synchrotron emission.})
and radio-loud AGN as a function of the 1.4-GHz flux density. A value of
$q_{24} = \langle q_{24}\rangle-2\sigma_{q_{24}}$ (i.e.\ 2\,$\sigma$ below
the mean value) is adopted as the cut-off value of $q_{24}$ used to
separate and distinguish `star-forming dominated' sources from radio-loud
AGN. In particular, the cut-off used in the main panel of
Figure~\ref{fraction24um} is at $q_{24}=-0.23$, which in terms of
luminosity means that radio-loud AGN are selected by having a radio power
$\sim9$ times or larger than that predicted by the mean value.  All radio
sources with measured 24-$\mu$m flux densities and $q_{24}$ values above
the threshold are indicated in Figure~\ref{fraction24um} as `Star-forming
Dominated'. The so-called `Radio-excess' fraction corresponds to all those
sources detected at 24\,$\mu$m with $q_{24}$ below the cut-off value, or
undetected but with upper limits below the threshold. The remaining
population of sources will be those that are undetected at 24$\mu$m but
have upper limits in $q_{24}$ that lie above the $q_{24}$ threshold for
AGN selection. Those will be a mixture of starburst/radio-quiet systems
and radio-loud AGN.

We have already said in Section~\ref{irrcorr} that the faint radio
population ($\rm \lesssim 300\,\mu Jy$) is affected by incompleteness
at 24\,$\mu$m. In fact, for sources that follow the FIR/radio
correlation then, using the threshold for detection in the SWIRE data,
we can predict the expected fraction of detections at 24\,$\mu$m as a
function of radio flux density. We can then use this to correct the
observed fraction of star-forming sources for incompleteness. In order
to do so, we ran Monte-Carlo simulations for each radio flux density
bin (from Figure~\ref{fraction24um}), using the mean and scatter of
the ${\rm 24\mu m/1.4GHz}$ correlation, to obtain the
probability of detection based on the 24-$\mu$m image threshold at
$200\,\mu$Jy. The correction for incompleteness to the ``Star-forming
Dominated'' population is called, in Figure~\ref{fraction24um}, the
``Estimated Star-forming Dominated'' fraction, which is mainly introduced
at faint radio fluxes. The reminder of the uncertain population is
classified as the ``Probable Radio-excess'' fraction.

The main panel of Figure~\ref{fraction24um} shows our best prediction
for the transition from radio-loud-AGN-dominated to a
star-forming-dominated population at sub-mJy radio fluxes. It makes
use of the mean and scatter of the M82-like $k$-corrected $q_{24}$
values (from \S3.3) to define the threshold value
and the inputs to the Monte Carlo simulations. It is found that for
star-forming sources with $S_{\rm   1.4GHz}=\rm 200-500\,\mu Jy$ ($\rm
10^{-3.7}- 10^{-3.3}\,Jy$), or with higher radio fluxes, the expected
fraction of counterparts at 24\,$\mu$m starts to become
complete. Actually, the Monte-Carlo estimation predicts that
$\gtrsim85$ per cent of the sources should have counterparts at
24\,$\mu$m in this bin (it changes to $\gtrsim98$ per cent considering
the Appleton et al.\ \citealt{Appleton04} $q_{24}$ values). This
confirms the previous estimation for the complete star-forming radio
sample based on a cut-off at $S_{\rm 1.4\,GHz}={\rm 300\, \mu Jy}$.

In order to show how the results change with uncertainties in
the FIR/radio correlation, the two small panels at the bottom in
Figure~\ref{fraction24um} use an exaggerated scattering for the
correlation, and the Appleton et al.\ (\citealt{Appleton04}) results
(left and right hand side, respectively), as indicated by the values
above the panels. Each case, however, keeps almost the same
fit to the high end of the $q_{24}$ distribution ($\langle q_{24}
\rangle+\sigma_{q_{24}}\approx1.2$).  

From our best fit we predict a significant
($25\pm5$ per cent) fraction of radio-loud AGN at flux levels of $S_{\rm
1.4GHz}\sim\rm 300\,\mu Jy$. The number of radio-loud AGN clearly
increases as a function of the radio flux density and above
$S_{\rm 1.4GHz}\gtrsim\rm 3.2\,mJy$ ($\rm 10^{-2.5}\,Jy$) essentially
all radio sources are radio-loud AGN. At
$\sim50\mu$Jy flux density levels, the fraction of radio-loud AGN
significantly decreases to zero ($\lesssim20\%$ from the lower panel
estimations), leaving the star-forming dominated systems as the only
population at this faint radio flux regime. 

These simple analyses suggest that a large number of radio-loud AGN
could be selected using deep mid-IR and radio surveys based on a
simple radio-excess selection criteria. In terms of the AGN
population, we note that our predictions for the ``Radio-excess''
fraction is just part of the total number of AGN, because many
radio-quiet AGN may be included in the `star-forming-dominated' population. 

\subsection{Host galaxy properties}

\begin{figure} 
  \includegraphics[scale=0.35]{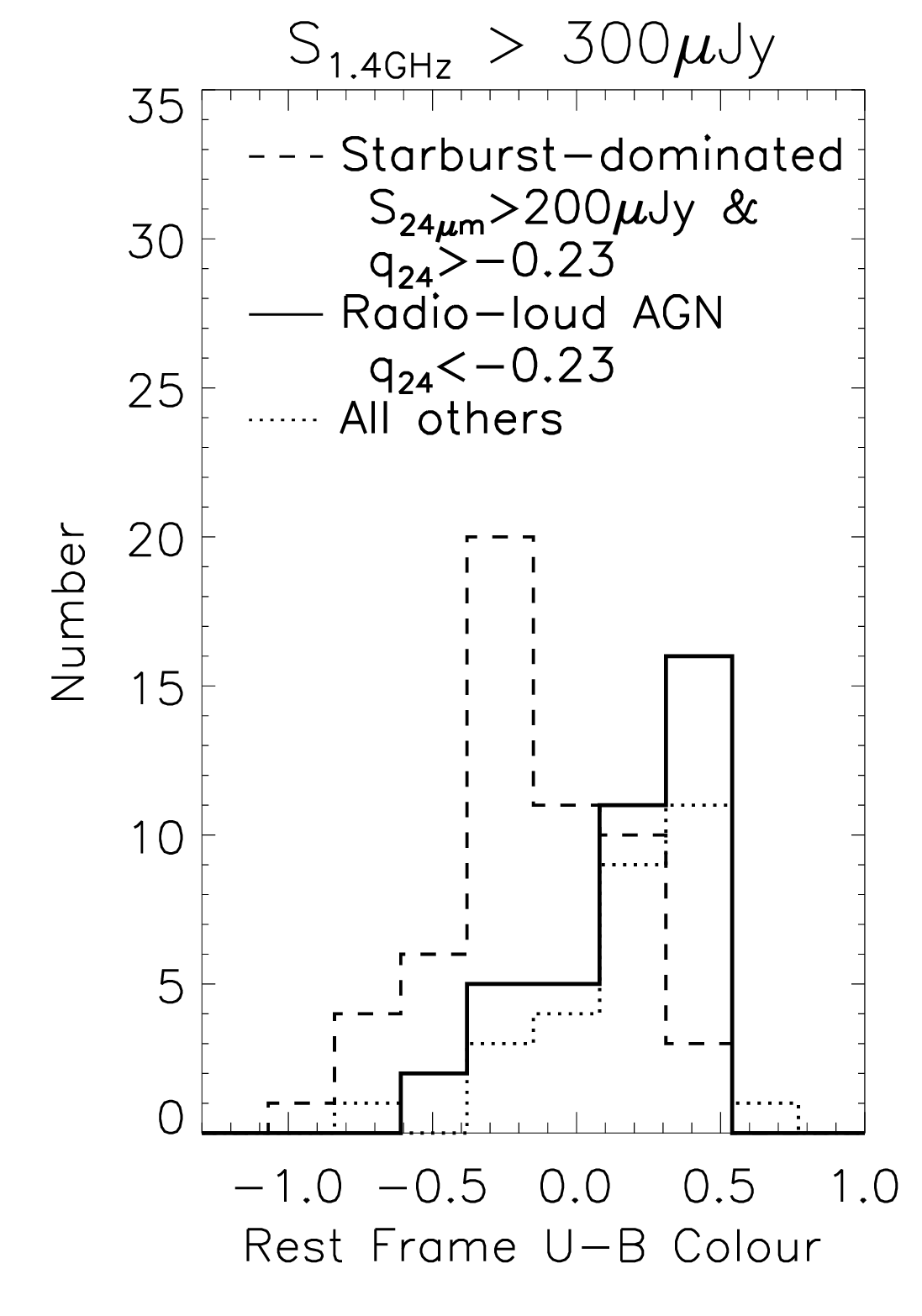} 
  \includegraphics[scale=0.35]{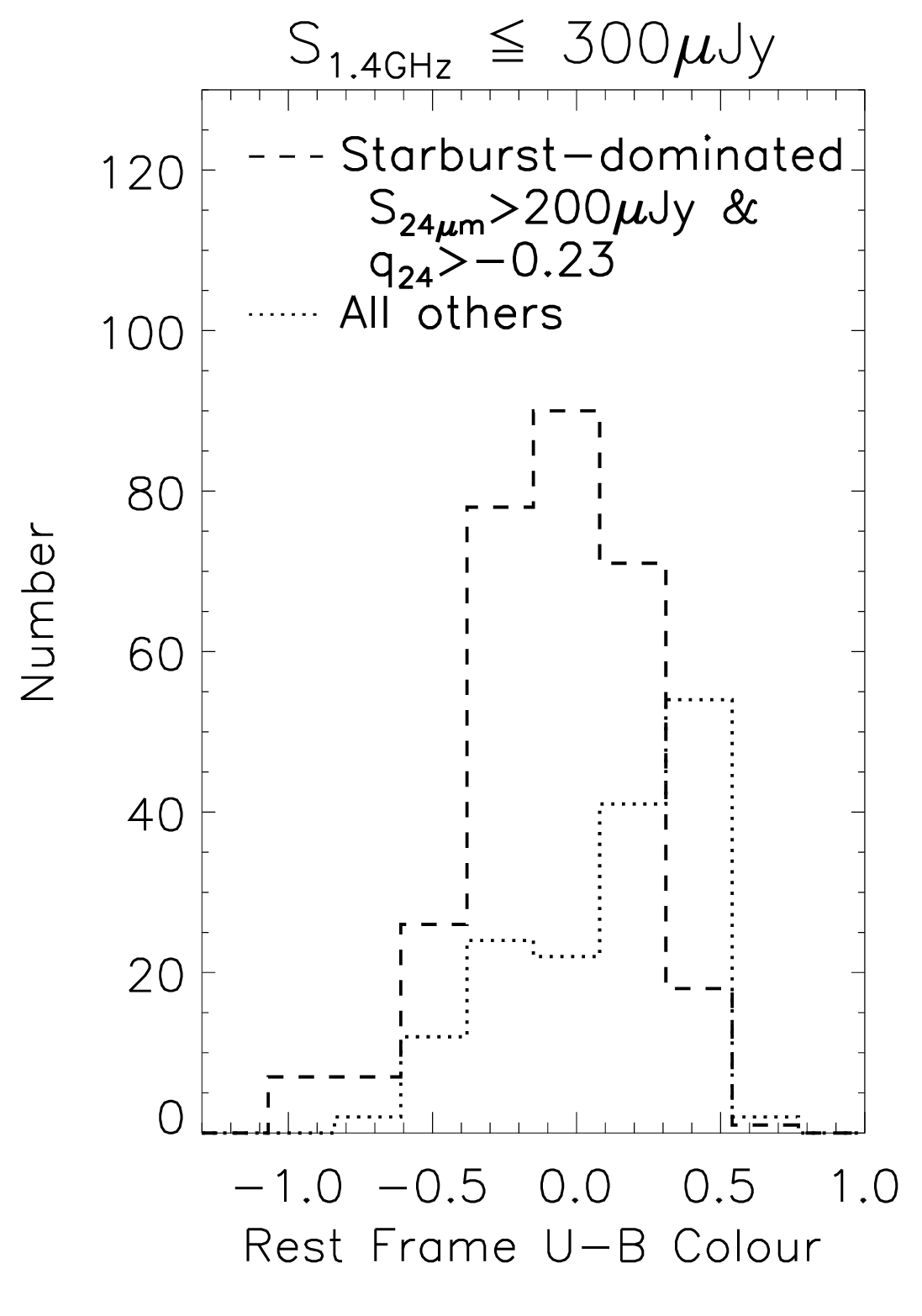} 
  \caption{ 
    Histogram of rest-frame $U$-$B$ colours of all sources with
    estimated redshift in two bins of radio density flux. Dashed,
    continuum, and dotted lines show the distribution of the
    ``Star-forming Dominated'', ``Radio-excess'' and ``All others''
    (``Estimated Star-forming'' and ``Estimated Radio-excess'')
    fractions, respectively. These correspond to the same ones plotted
    in the main panel of Figure~\ref{fraction24um}.
    It shows that the Radio-Excess population has a fundamentally
    different nature to the Star-forming Dominated sources, and also 
    confirms that at bright radio flux densities ($>300\mu$Jy) the 
    ``All Others'' population is comprised almost entirely of 
    radio-excess objects.
    }
  \label{colours} 
\end{figure}

To explore the optical galaxy host properties of the
different populations presented in \S4.1, we estimated
the rest-frame ($U$-$B$) colour (limited to sources with
estimated redshift) for each source. The colour histograms for
the three populations, ``Star-forming Dominated'' sources, the
``Radio-excess'' sources, and ``All others'' (``Estimated Radio-excess'' 
and ``Estimated Star-forming Dominated'' fractions combined), are 
plotted in Figure~\ref{colours}.

Considering only the population with radio flux densities $S_{\rm
  1.4GHz}>\rm 300\,\mu Jy$ (left-hand panel), we find that both the
``Radio-Excess'' and ``All others'' populations present clearly redder
distributions compared to the the ``Star-forming Dominated'' population. A
Kolmogorov-Smirnov (KS) rejects the hypotheses that either the
``Radio-Excess'' or ``All others'' sources are drawn from the same parent
population as the ``Star-forming Dominated'' sources at $>99.999$\%
significance in both cases, proving that they have a different nature.  On
the other hand, the colour distributions of the ``Radio-Excess'' and ``All
others'' populations are statistically indistinguishable. This
similarity is expected because, as argued in \S 3.3, at these radio
flux densities 
the ``All others'' population ought to be composed essentially entirely of
``Radio-Excess'' sources (see Figure~\ref{fraction24um}). At lower radio
flux densities, the ``All others'' population should contain a mixture of
starburst-dominated galaxies and radio-loud AGN, and the right-hand panel
of Figure~\ref{colours} shows that indeed at these lower radio flux densities
this population has a more pronounced tail towards bluer $U-B$ colours.

The red colours found for the radio-excess sources may come from the
absence of star formation or due to dust obscuration in these
sources. Since they do not have detections at 24\,$\mu$m -- probable
absence of re-emission -- we suggest that this radio-loud AGN population is
hosted in red-galaxies with little or no star-formation. This is
potentially a very useful way to detect obscured AGN missed in deep X-Ray
surveys, which are responsible of the bulk of the Cosmic X-Ray Background
(Ueda et al. \citealt{Ueda03}).

\section{Discussion} 
\label{discussion} 

Previous studies of the universality of the FIR/radio correlation at high
redshift have combined 1.4-GHz data with {\it ISO} 15-$\mu$m data (Garret
\citealt{Garret02}) and {\it Spitzer} 24- and 70-$\mu$m data (Appleton et
al.\ \citealt{Appleton04}), both showing evidence of a correlation up to
$z\sim\rm 1$. More recently, Kov\'acs et al.\ (\citealt{Kovacs06}) and
Beelen et al.\ (\citealt{Beelen06}) have shown evidence for the
correlation at higher redshifts, for a sample of 15 submm galaxies
($z\sim\rm 0.5-3.5$) and for a handful of distant quasars ($z\sim\rm
2-6$), respectively.  In this work, despite the difficulty of measuring a
reliable bolometric FIR emission, the monochromatic 24-$\mu$m flux
densities show a tight correlation with the 1.4-GHz flux density up to
$z\approx3.5$.  $k$-corrections based on an M82-like template indicate
that both the mean and the scatter of the $q_{24}$ values remain constant
as a function of redshift (see Figure~\ref{evolution}).

The mean and scatter of the distribution of $q_{24}$ values obtained after
a M82-like $k$-correction for all radio sources detected at 24\,$\mu$m is
$q_{24} = 0.99\pm0.32$. This is in broad agreement with the results of
Appleton et al.\ (\citealt{Appleton04}), $q_{24}=0.84\pm0.28$, based
on a sample $z\lesssim1$. However, if our sample is restricted to
brighter radio sources for which all star-forming sources would be
expected to have been detected at 24\,$\mu$m, then the mean relation
becomes $q_{24} = 0.71\pm0.47$, with a 
scatter about twice that of Appleton et al. The median error value for the
$q_{24}$'s has been found to be about 0.1 dex, which suggests that this
scatter is intrinsic for the correlation and does not come from
observational errors. We can conceive of two possible explanations for
this increased scatter at higher radio flux densities. First, it may be
that, because of the smaller number of sources, some radio-excess objects
may fail to be excluded by the biweight estimator, and therefore both
bias and increase the scatter of the relation. Second, it may be that for
these more powerful sources, the M82-like template (a relatively low
luminosity source) is not an accurate representation of the FIR emission
of all sources, but that a broader range of SED types is present,
increasing the scatter.  Figure~\ref{q24_z_kcorr} provides some evidence
for this: although a very flat $q_{24}$ distribution is found in the full
radio sample (top panel), some redshift variation may be present in the
brighter radio sources (lower panel), in particular the presence of a
handful of sources with higher $q_{24}$ values at low redshifts $z<0.5$,
which would be more consistent with a steeper mid-IR slope, such as that
given by the Arp220-like template.

Recently, Boyle et al.\ (\citealt{Boyle07}) have found a significantly
higher mean value, $q_{24}=1.39\pm0.02$ ($2\,\sigma$ away from our
estimation) whilst applying stacking techniques (Wals et al.\ 
\citealt{Wals05}) to ATCA 1.4-GHz images using positions from
{\it{Spitzer}} detections in the CDFS and ELAIS fields. Their work
lacks redshift information, however, and therefore does not include
$k$-corrections. We have shown the relevance of different mid-IR
features for high redshift sources at the flux regimes of this work,
suggesting a clear underestimation for the intrinsic $24\mu$m/1.4GHz
scatter in their work. Also, a stacking analysis for a mixture of 
star-forming galaxies and radio-loud AGN does not discriminate an
exclusive correlation for star-forming galaxies. Since the flux
thresholds from our work are similar to their ones, we suggest that
selection effects based on different sample selection criteria (from
radio or mid-IR detections) may also be affecting the scatter on the
24$\mu$m/1.4GHz correlation. 

The FIR/radio correlation in not precisely linear, specially in
optically selected samples containing faint galaxies.
It is expected that the radio emission, in low luminosity
galaxies, is more deficient since cosmic rays are more likely to
escape from the galaxy (Chi \& Worfendale
\citealt{Chi90}). At redshifts larger than unity the Cosmic
Microwave Background might quench the radio 
emission (Condon \citealt{Condon92}) and a cosmic evolution for
the magnetic fields cannot be rejected too. 
Nevertheless, in this work the exploration of more detailed
models, showing for example luminosity (or mass) dependencies for
the $q_{24}$ parameter, was limited by the uncertainties in
photometric redshifts.

A simple radio flux estimation ($S_{\rm 1.4GHz}^{limit}\approx\rm
35\,\mu Jy$) shows that a starburst galaxy with a typical M82
luminosity can be observed up to moderate redshifts only, $z\sim\rm
0.4$. In fact, it is expected that the observed high-redshift
high-luminosity Universe is composed mainly of ULIRGs and AGN. 
ULIRGs are usually related to interacting systems (Sanders et al.\
\citealt{Sanders88}) and are expected to suffer strong absorption,
implying large $k$-corrections for the 24-$\mu$m emission. 
Insights for steeper mid-IR SED at $S_{\rm 1.4GHz}> {\rm 300\mu
  Jy}$ are hinted (bottom Figure~\ref{q24_z_kcorr}), however, the lack
of a clear scatter excess in the FIR/radio distribution at
$z\approx\rm 1-2$ (Figure~\ref{q24_z}) suggests that the bulk of the
ULIRG-like high-redshift sources from our sample do not suffer large
variations due to silicate absorption. The correlation found in this
work is therefore quite unexpected, but suggests that a M82-like
mid-IR template, intrinsically coming from a lower luminosity source,
is still a good representation for the mid-IR SED of high-redshift
star-forming-dominated systems. The M82-like template is also
supported by Lagache's ULIRG ($L_{IR}=10^{12}M_\odot$) model, and
particularly leaves Arp\,220 as single extreme case.
These results are more relevant considering that Kasliwal et al.\
(\citealt{Kasliwal05}) estimate that about half of 
ULIRGs at $z\approx\rm 1-1.8$ may be unobserved at 24\,$\mu$m due to
silicate absorption at 9.7\,$\mu$m in the Bootes field, a suggestion
not particularly supported by our data. For ULIRGs selected in sub-mm
observations, their mid-IR SED is still under debate, but templates
such as M82 and Arp\,220 (Menendez-Delmestre et
al. \citealt{Menendez07} and Pope et al.\ \citealt{Pope06},
respectively), have been found as good fits for their mid-IR range. 

In Section~\ref{evidence_agn}, we have shown the capabilities of deep
mid-IR and radio imaging for the selection of radio-loud AGN
(``radio-excess'') based on different cut-offs in $q_{24}$.  We have
shown that at $S_{\rm 1.4GHz}>\rm 300\,\mu Jy$, selected radio-excess
sources strongly suggests nuclear activity. The radio-excess
sources present a redder distribution of rest-frame $U$-$B$ colours
compared with those identified as star-forming-dominated
galaxies. This strongly suggest a different nature between the
populations selected by the simple $q_{24}$ threshold at -0.23.
Previous population synthesis models (Jarvis \& Rawlings
\citealt{Jarvis04}) have predicted that radio-excess sources at these
flux densities are mainly FR\,{\sc I} (Fanaroff \& Riley
\citealt{Fanaroff74}). It is known that FR\,{\sc I} do not fall
into the standard unified scheme for AGN (Antonucci
\citealt{Antonucci93}) and show no bright AGN nucleus. Therefore their 
red colours indicate that these radio-loud sources do not present 
significant star-formation
activity and/or are hosted in red massive galaxies. 

Jarvis \& Rawlings (\citealt{Jarvis04}) predict that approximately
half of the sources with radio fluxes between  ${\rm 100\,\mu Jy} \leq
S_{\rm 1.4GHz}< {\rm 300\,\mu  Jy}$ present radio-loud activity. We
have predicted from our best fit a fraction of $25\,\pm\,5\,\%$ at
this flux regime, in disagreement by a factor of two with their
estimation.  Figure~\ref{fraction24um} shows that Appleton et
al. estimation may explain a 50\% fraction, although we demonstrated
in \S3.3 that incompleteness biases the assumptions of those results.
On the other hand, Simpson et al. (\citealt{Simpson06}) 
tentatively suggested a radio-quiet AGN fraction of 20\,$\%$ at this
radio fluxes regime, but since radio-quiet sources have been found
to also follow the FIR/radio correlation, in this work we are not able
to discriminate between this population and starburst
galaxies. Therefore, this radio-quiet AGN population remains
uncertain. 

\section{Conclusions}
\label{conclusion}

We have analysed the 24\,$\mu$m properties of a radio-selected sample
in order to explore the well-known FIR/radio correlation at high
redshifts. We have shown evidence that 24-$\mu$m data do not give the
best estimation of the total bolometric FIR emission, reflected in a
considerably larger scatter on the correlation than previous work
using longer wavelengths ({\em IRAS}/60\,$\mu$m -- Yun et al.\
\citealt{Yun01}, {\it Spitzer}/70\,$\mu$m -- Appleton et al.\
\citealt{Appleton04}). Nevertheless, despite the large scatter, the
IR-24\,$\mu$m/radio-1.4\,GHz correlation seems to extend up to
redshift $\sim3.5$ with roughly the same mean and scatter. The fact
that we do not observe considerable variations in the correlation as a
function of redshift, mainly seen on the well defined upper end of the
$k$-corrected $q_{24}$ distribution (see top
Figure~\ref{q24_z_kcorr} \& \ref{evolution}), highly suggests its
validity all the way back to primeval times. This is the first evidence
for the extension of the correlation over this very wide range of
redshifts. 

Statistically, the correlation is described by the basic observable
parameter $q_{24}$, the ratio between the flux densities at 24\,$\mu$m
and 1.4\,GHz, respectively. The $q_{24}$ factor parameterise the link
(in the local Universe at least)
that star-forming galaxies have between the synchrotron emission, from
cosmic rays accelerated by massive supernovae explosions, i.e. the
massive end of the mass function in galaxies, with the reprocessed UV
photons re-emitted by the dust at FIR wavelengths. Besides an
intrinsic scatter for the correlation, the
presence of an AGN may also modify the mid-IR spectra and the
radio emission, biasing, and therefore increasing the scatter on a 
$q_{24}$-based correlation. We found a mean value (biweight estimator)
of $q_{24}=0.30\pm0.56$ for the observed and $q_{24}=0.71\pm0.47$ for
the $k$-corrected (M82-like) data based on an expected complete radio
sample ($S_{\rm 1.4GHz}>\rm 300\,\mu Jy$) at 24\,$\mu$m.

Three different $k$-corrections have been tested: two based on
well-known local starburst sources (M82 and Arp\,220) and one on a SED
ULIRG model (Lagache et al.\ \citealt{Lagache04}). The main difference
introduced by using different templates is caused by the silicate
absorption feature at $\sim$10\,$\mu$m being redshifted into the
observed 24-$\mu$m band at $z\approx1-2$ (Figure~\ref{q24_z}). 
Based on the smallest scatter found for $q_{24}$ data, we propose an
acceptable $k$-correction given by an M82-like mid-IR template,
which suggests no strong evidence for extreme silicate absorption in the
bulk of our sample.
We do, however, find tentative evidence for a 
steeper mid-IR spectrum than M82 for brighter radio sources 
$>300\mu$Jy, in order to explain the slight offset from the correlation 
for sources with $z<0.5$.

We have found a clear dependency for the detected number of
star-forming-dominated galaxies (starburst and radio-quiet AGN) as a
function of the radio flux density. Making use of the derived
$\rm 24\mu m/1.4GHz$ correlation, and the flux density threshold from the
24-$\mu$m image, we ran Monte-Carlo simulations to estimate the
incompleteness at 24\,$\mu$m, and therefore the intrinsic fraction of
star-forming-dominated galaxies. This allowed us to estimate the
probable fraction of radio-loud AGN and to describe the transition from
AGN-dominated to star-forming-dominated galaxies at faint radio flux
densities ($\lesssim1\,\rm mJy$). Our best-fit estimation predicts a significant
fraction ($25\,\pm\,5$ per cent) of radio-loud AGN at
$S_{\rm 1.4GHz}\sim\rm 300\,\mu Jy$, rising quickly to become a dominant 
AGN population at brighter radio flux densities ($>1\,\rm mJy$), and
falling towards zero (certainly $\lesssim20$ per cent) at fainter fluxes. 
These results are in broad agreement with previous
studies (e.g.\ Condon \citealt{Condon84}).

We have also shown that the rest-frame $U$-$B$ colours, of the sources
that are expected to be radio-loud AGN, have a redder distribution than
the starburst/radio-quiet AGN galaxies (Figure~\ref{colours}),
strongly suggesting a different nature between the sources selected by
our simple $q_{24}=-0.23$ threshold. These radio-excess sources seem to be 
hosted in red galaxies with little or no star-formation implied by the
non-detection at $24\mu$m.  This is therefore an extremely promising way
to select obscured Type-2 AGN (Comastri et al.\ \citealt{Comastri03}),
with a radio-loud nature,
which are missed by deep X-ray observations. Nevertheless, future
spectroscopic follow-up of these sources, and also a better description of
the FIR emission, is needed to fully test this method.

\section*{Acknowledgements} 
This paper was supported by a Gemini--STFC research
studentship. E.I. thanks to A. Pope, G. Lagache, P. Lira \&
B. Brandl for templates and useful comments to the paper. P.N.B.,
R.M., I.R.S. and O.A. acknowledge support from the Royal Society.

\label{lastpage} 
 
\end{document}